\documentclass{sig-alternate}
\usepackage{times}
\usepackage{graphicx}

\makeatletter
\newif\if@restonecol
\makeatother

\usepackage[ruled, vlined]{algorithm2e}
\usepackage{listings}

\begin{document}

\date{}

\title{A Metamodel of Unit Testing for Object-Oriented Programming Languages}

\author{
Martin L\'evesque\\
Dept. of Computer Science\\
University of Quebec at Montr\'eal\\	
Montr\'eal, Qu\'ebec, Canada\\
Email: levesque.martin@gmail.com \\
}

\maketitle
\thispagestyle{empty}

\begin{abstract}
{A unit test is a method for verifying the accuracy and the proper functioning of a portion of a program. This work consists to study the relation and the approaches to test \textit{Object-Oriented Programming} (OOP) programs and to propose a metamodel that enables the programmer to write the tests while writing the source code to be tested.} \end{abstract}




\section{Introduction}

\lstset{language=Java}
\lstset{frame=single}
\lstset{captionpos=b}
\lstset{frameround=fttt}
\lstset{numbers=left}
\lstset{numberstyle=\tiny}
\lstset{tabsize=2}
\lstset{basicstyle=\footnotesize}

	A unit test \cite{article:unitTestSurvey} is a method for verifying the accuracy and the proper functioning of a portion of a program. This work consists to study the relation and the approaches to test OOP programs and to propose metamodel that enables the programmer to write the tests while writing the source code to be tested. By considering the current approaches used in the industry, the integration and the writing of the tests are hard tasks: most likely, the tools available require to write additional components manually and most of the time, the tests are not near of the source code to be tested.

	This paper is organized as follows. Section \ref{sec:OOP_unit_tests} gives a state of the art about the relation between OOP programs and the tests. Section \ref{sec:metamodel} describes the proposed metamodel of unit testing for OOP languages. Section \ref{sec:example} gives an example of what could looks like an OOP environment which uses the proposed metamodel. finally, Section \ref{sec:conclusion} contains the conclusion and future possible work.

\section{OOP and Unit Tests}
\label{sec:OOP_unit_tests}

	This section contains the motivations of testing OOP languages differently compared to testing procedural languages. Then, the levels of abtraction for the tests are defined. Finally, current approaches to test OOP languages are discussed.

\subsection{Motivations of testing OOP languages differently compared to testing procedural languages}

	This section describes some reasons why an OOP language must be tested completely differently compared to procedural languages.

	Intuitively, specialization and aggregation combined to polymorphism increases the difficulty to detect errors during the integration of several components. The primary goal consists of decreasing the maximum effort to test a unit by reusing as much as possible the tests \cite{article:ISE-TR-99-05, article:OOTestingReport}.

	\subsubsection{Test unit}
	
	In procedural languages such as (C, Pascal, etc.), the principal units are the procedure and the module. However, the principal unit in an OOP language is the class \cite{article:ISE-TR-99-05, article:openIssues, article:issuesTestingOOSoft}. This difference has a huge impact on the methodology to test the software since several new OOP concepts are nonexistent in the procedural paradigm.

	\subsubsection{Object and encapsulation}
	
	Traditional techniques to test procedural languages are not necessary application to the object oriented paradigm since the encapsulation reduces or eliminates the possibility to test a specific state of an attribute of an object \cite{article:ISE-TR-99-05}. 

	\subsubsection{Specialization and inheritance}
	
	A method is defines by one of the following characteristic \cite{article:ISE-TR-99-05}:

\begin{enumerate}
\item Inherited from the parent class without redefinition.
\item Redefinition without any call to super.
\item Redefinition with a call to super.
\item A new method nonexistent in the parent classes.
\end{enumerate}

	Existing works consider that the tests for cases 2, 3 and 4 must be retested completely compared to the first case which does not require any redefinition. 

	\subsubsection{Polymorphism}
	
	Let $p(z)$ be a method named $p$ which take a parameter $z$ where $z$ contains a method $m$. Depending on the dynamic type of $z$, several different methods $m$ when $z.m$ is called by polymorphism. It has a huge impact on the tests since several permutations of dynamic types must be passed for the parameters in order cover more lines of code.

\subsubsection{OOP language entity dimensions}

	OOP language entities can be categorized as follows \cite{article:openIssues}:

\begin{itemize}
\item Object: A class instance composed of several attributes and methods. An object represents a specific state.
\item Class: A model which factorizes the properties of its instances.
\item Class hierarchy: A class inherits the properties of its parent classes.
\item Package (or subsystem): A subsystem defines an interface presenting services. A subsystem regroups classes with strong semantic relations. Jin et al. has developed a mechanism which uses a metamodel which allows to test the coupling between units \cite{article:couplingBased}.
\item System: Composed of a set of subsystems and links them together.
\end{itemize}

\subsection{Levels of abstraction for the tests}

	The principal levels of abstraction for the test are the white box, the black box and the grey box \cite{article:issuesTestingOOSoft}:

\begin{itemize}
\item White box: White box tests consider the implementation of the classes and the methods.
\item Black box: Black box tests does not take into account the implementation but only the interface units.
\item Grey box: Grey box tests take into account the white box and the black box levels.
\end{itemize}

\subsection{Current approaches to test OOP languages}
	
	The approaches used in practice and in the research community are based on specifications or on programs.

\subsubsection{Specification-based}

	The first approach to test an object-oriented program consists to use a certain specification language which describes what the packages and the classes are actually doing. With this specification, it is possible to generate several tests automatically in order to verify the correctness of the implementation.

	Barbey et al. \cite{article:theorySpecBasedOOSoft} have used an object oriented specification called \textit{Concurrent Object-Oriented Petri Nets} (CO-OPN/2). In such a language, an interface describing the attributes in the methods is defined. The axioms enable the programmer to define the comportements of the methods. The main problem in this work is that it does not take into account the characteristics of the OOP paradigm such as polymorphism, specialization, etc.

	Several researchers \cite{article:astoot, article:automaticIdUnitTests} have defined that the state of an object is the result of a certain number of method calls. Doong et al. \cite{article:astoot} have used a similar approach compared to the work of Barbey and al., however they specified algebraically the behavior of the classes. The verification of the correctness of a method is done by generating a sequence of method calls. Then axioms are used to validate the results.

	The main problems with the specification-based approach is that the programmer must manipulate an extra language only in order to test the software. Also, the tests are most likely not near the source code to be tested which is not good by considerating that, in practice, the tests are not updated when the software change.

\subsubsection{Program-based}

	The program-based approach consists to get a certain confidence level by writing manually the tests in order to simulate the main possibilities of execution of the software.

\begin{figure}
\centering
\includegraphics[scale=0.45]{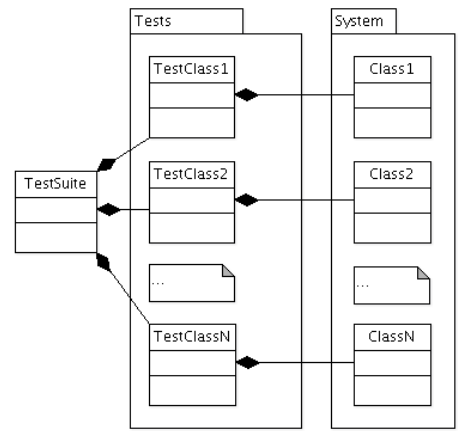}
\caption{System of N classes uing JUnit.}
\label{fig:junit_system}
\end{figure}

	A standard in the industry is the \textit{JUnit} Framework \cite{article:jUnit} well used for the Java programming language. The main components of the \textit{JUnit} Framework are the tests suites and the tests cases. A test case regroups several tests of a class and the suite is used to contains several tests cases. Let say a system has $N$ classes. Then if there is one test case for each class, then the system will consists of $2 * N + 1$ classes (Fig. \ref{fig:junit_system}).

\begin{figure}
\centering
\includegraphics[scale=0.45]{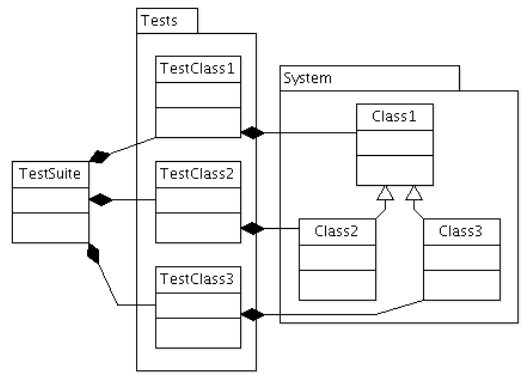}
\caption{Loss of the semantic relations with a typical system using JUnit.}
\label{fig:loss_semantic_relations}
\end{figure}
	Another problem with \textit{JUnit} is the loss of the semantic relations between test classes (Fig. \ref{fig:loss_semantic_relations}). In fact, classes \textit{TestClass2} and \textit{TestClass3} could be defined to be the subclass of TestClass1 and reuse the code from \textit{TestClass1}. However it would be done manually. We need something more automatic in order to reuse the tests. 

\section{Metamodel of Unit Testing for OOP}
\label{sec:metamodel}

\begin{figure}
\centering
\includegraphics[scale=0.35]{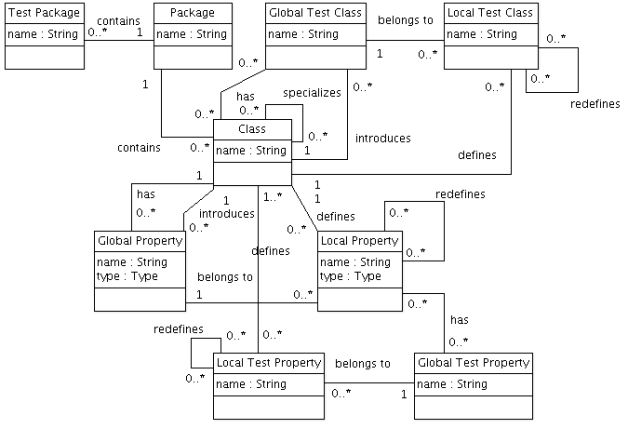}
\caption{Metamodel of Unit Testing for OOP.}
\label{fig:modelePropose}
\end{figure}

	The proposed metamodel (Fig. \ref{fig:modelePropose}) is an extension of the metamodel defined by Ducournau et al. \cite{article:metaModelUML}. The main motivation of this extension is to add properties specifically for unit tests in order to write unit tests by taking advantage of the mechanisms and the forces of OOP.

	A test property (\textit{Test Package}, \textit{Local Test Class} or \textit{Local Test Property}) represents, in a complete OOP environment using the proposed metalmodel, a block of unit test code associated to a unit (A Package, a class or a property).

	The idea of incorporating test properties inside a class in the metamodel was inspired by the D Programming Language \cite{site:dProgramming}. However, they do not considered OOP mechanisms such as inheritance and they have not modeled several layers of granularity.

	Here is a brief list of the main features of the proposed metamodel:

\begin{itemize}
\item Covered dimensions: The proposed model enables to test packages (and sub-packages), classes and class properties.
\item Level of abstraction to define tests: Grey box. Testing packages allows to link classes together regardless of the implementation, which corresponds to the black box level. However, testing classes and methods corresponds to the white box level since the tests and the implementation are regrouped within the same entity.
\item The program-based approach to test the software is used.
\item Test's units are the package, the class and the property class. On the other hand, the central unit is the class.
\item Encapsulation: the white box tests allow to access to all properties compared to the black box tests which allow to access only to public properties.
\item Inheritance and specialization of tests are built by considering the real world and the type safety. A super class' test is useful to test a subclass, so it is inherited by default.
\item The specialization is used for the tests class and for the tests property in order to reuse
\end{itemize}

\subsection{Class and properties}

	As said, the proposed metamodel (Fig. \ref{fig:modelePropose}) is an extension of the metalmodel proposed by Ducournau et al. \cite{article:metaModelUML}. However, properties for tests are the following:

\begin{itemize}
\item Properties to test classes.
\item Properties to test class properties.
\item Properties to test packages.
\end{itemize}

	The relations defined in \cite{article:metaModelUML} are also used and expressed as follows:

\begin{itemize}
\item \textit{Redefines (redef)}: Used to redefines a given entity.
\item \textit{Has (has)}: Used to know which knowledge is known by a certain entity. For example, it can be useful to know which tests are attached to a given local property.
\item \textit{Introduces (intro)}: Useful to know which entity has introduced a property or a test.
\item \textit{Belongs to (belongs)}: Useful to know which global entity is associated to a given local entity.
\item \textit{Defines (def)}: Used to know which entity has defined a certain property.
\end{itemize}

\subsubsection{Property tests}

\paragraph{Global Test Property}

	Let's express the global test properties which belongs to a certain global property ($g_c$) of a class $c$ having super classes. $Parents_c$ represents the set of all super classes of $c$:

\begin{equation}
\label{formule:parents}
	\forall_{p \in Parents_c} c \prec p
\end{equation}

	Let $G_c$ be the global properties of $c$, $L_{g_c}$ the local properties of a given global property $g_c \in G_c$ of a class $c$ and its super classes are defined by:

\begin{equation}
\label{formule:proprietesLocales}
	L_{g_c} = \bigcup_{d \in Parents_c \cup \{c\}} \{l_d | belongs(l_d, g_c), l_d \in L_d\}
\end{equation}

	where \textit{belongs} is a relation used to know if a local property belongs to a global property.

	Then the global test properties ($GTP_{g_c}$) of a given global global property is given by:

\begin{equation}
\label{formule:testsGlobauxProprieteGlobale}
	GTP_{g_c} = \bigcup_{l_c \in L_{g_c}} \{gtp | has(l_c, gtp), gtp \in GTP_{c}\}
\end{equation}

	where $GPT_c$ represents the global test properties of a class $c$. The relation $has(l_c, gtp)$ states that $l_c$ knows $gtp$.

	A \textit{Global Test Property} (GTP) is either inherited from a local property redefinition or introduced by a local property. The introduced global tests ($IGTP_{g_c}$) represent a sub set of $GTP_{g_c}$:

\begin{eqnarray}
\label{formule:testsGlobauxProprieteGlobale}
	\bigcup_{d \in Parents_c \cup \{c\}} & & \{ gtp | intro(d, gtp), \nonumber \\ 
	& & \exists_{l_c \in L_{g_c}} has(l_c, gtp), \nonumber \\ \nonumber \\
	& & gtp \in GTP_{c}\} \subseteq GTP_{g_c}
\end{eqnarray}

	where $intro(d, gtp)$ means that $d$ introduces $gtp$.

\paragraph{Local Test Property}

	Let define all local test properties $LTP_c$ for a given class $c$, the local test properties of a given global property $gt_c$ is defined by:

\begin{equation}
\label{formule:testsLocauxTestGlobal}
	LTPC_{gt_c} = \bigcup_{lt_c \in LTP_c} \{ lt_c | belongs(lt_c, gt_c)\}
\end{equation}

	However, redefined tests must be removed:

\begin{eqnarray}
\label{formule:testsLocauxTestGlobalSansRedefinis}
	LTP_{gt_c} & = & LTPC_{gt_c} - \nonumber \\
				& & \bigcup_{t_1 \in LTPC_{gt_c}} \{ t_1 | \exists_{t_2 \in LTPC_{gt_c}} redef(t_2, t_1)\} \nonumber \\
	&	& 
\end{eqnarray}

	where $redef(t_2, t_1)$ is a relation which expresses that the test $t_2$ redefines the test $t_1$.

\subsubsection{Class tests}

\paragraph{Global Test Class}

	A global test class is either inherited from a super class or introduced by a given class. Let $GTC$ be the global tests class of a system, $GTC_c$ the global tests class of a class $c$ are defined by:

\begin{equation}
\label{formule:testsGlobauxClasse}
	GTC_c = \bigcup_{t \in GTC} \{ t | \exists_{c \in Parents_c \cup \{c\}} intro(c, t)\}
\end{equation}

\paragraph{Local Test Class}

	Let $LTC$ be the local tests of a class. The local tests of a class $c$ is given by:

\begin{equation}
\label{formule:testsLocauxClasseComplet}
	LTCC_{c} = \bigcup_{lt \in LTC} \{ lt | \exists_{gtc \in GTC_c} belongs(lt, gtc)\}
\end{equation}

	However, redefinitions must be removed:

\begin{eqnarray}
\label{formule:testsLocauxClasse}
	LTC_{c} = & LTCC_{c} - \nonumber \\
		& \bigcup_{t_1 \in LTCC_{c}} \{ t_1 | \exists_{t_2 \in LTCC_{c}} redef(t_2, t_1)\}
\end{eqnarray}

	\subsection{Type-Safety}
	
	In static typing languages, the redefinition of the return type must covariant and the redefinition of the parameters must be contravariant in order to have safe types. It does not poses any problem for the proposed metamodel since the metamodel is an extension of the existing properties and the notion of property and class is not altered.

	\subsection{Property Conflicts}

	\subsubsection{Global Test Property}
	\paragraph{Name conflict of tests with inheritance: by using two times the same global test property associated to the same local property of a super class}

\begin{figure}
\centering
\includegraphics[scale=0.5]{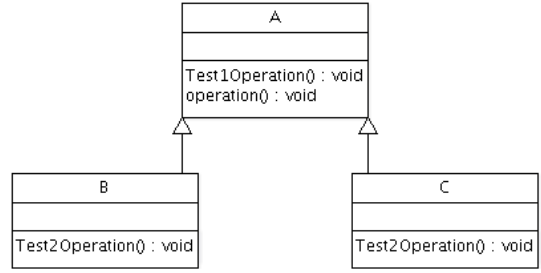}
\caption{Example of name conflict of tests with inheritance: by using two times the same global test property associated to the same local property of a super class.}
\label{fig:propertyConflict1}
\end{figure}

	The example in Fig. \ref{fig:propertyConflict1} is an example of a name conflict of tests by using two times the same global test property associated to the same local property of a super class. In this example, two tests for the same method \textit{operation} are defined (\textit{Test1Operation} and \textit{Test2Operation}). Both classes $B$ and $C$ define a new test \textit{Test2Operation} which causes a name conflict since two global test property with the same exists for the method \textit{operation}.

	Solutions:

	\begin{enumerate}
\item Factorize \textit{B.Test2Operation} and \textit{C.Test2Operation} in the super class $A$ if the test in class $B$ is the same than the one defined in class $C$.
\item Redefine \textit{operation} in $B$ and $C$.
\item Move \textit{Test2Operation} in class $A$ and redefine the local test property associated to \textit{Test2Operation} in class $B$ and $C$.
\end{enumerate} 

	\paragraph{Multiple Inheritance}

\begin{figure}
\centering
\includegraphics[scale=0.5]{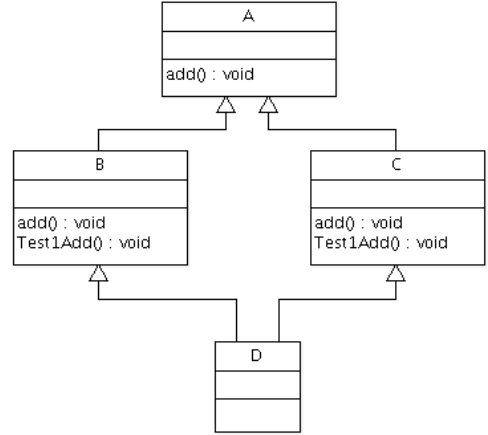}
\caption{Example of name conflict of global property test with multiple inheritance.}
\label{fig:propertyConflict2}
\end{figure}

	The second type of possible conflict with global test properties is with multiple inheritance. The example with multiple inheritance (Fig. \ref{fig:propertyConflict2}) does not cause any conflict for the tests since each new global test property is associated to the redefined \textit{add} method. However, the method \textit{add} is redefined in classes $B$ and $C$ and there exist three solutions for this problem. The tests must follow one of these three methods of conflict resolution:

	\begin{enumerate}
\item Rename method \textit{add} in $B$ and $C$ (for example \textit{B.addB}, \textit{C.addC}).
\item Selection. In class $D$, the method \textit{add} is redefined and a method between $B.add$ and $C.add$ is called and the associated tests of that selection are now associated also to $D.add$.
\item Unification. In class $D$, all tests from $B$ and $C$ are associated to $D.add$.
\end{enumerate}

\subsubsection{Local Test Property}

\paragraph{Local Test Property name conflict with inheritance}

\begin{figure}
\centering
\includegraphics[scale=0.5]{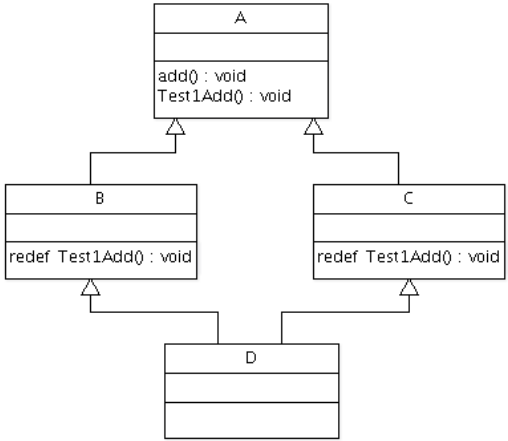}
\caption{Example of Local Test Property name conflict with inheritance.}
\label{fig:propertyConflict3}
\end{figure}

	Let $A$ be a super class of $B$ and $C$ which introduces a method \textit{add} and a test associated (Fig. \ref{fig:propertyConflict3}). The subclasses $B$ and $C$ redefine the local test property \textit{Test1Add}. Without class $D$, there exists no conflict. However, with a class $D$ which can inherit both \textit{B.Test1Add} and \textit{C.Test1Add} there is a name conflict: which test should be chosen by the compiler ? Depending on the test, three strategies exists with multiple inheritance: rename, select and unify.

	\subsection{Tests execution}

	A compiler of an object oriented programming language can then execute automatically by using the meta information from the metamodel proposed (Algorithm \ref{alg:executionTests}).

\begin{algorithm}[H]
\SetLine
\linesnumbered

\KwData{$P$, a set of all packages.}

\ForEach{$p \in P$} {
	// Test all packages

	\ForEach{Test Package \textbf{tp} such that $has(p, tp)$} {
		Execute $tp$
	}

	\ForEach{Class \textbf{c} such that $has(p, c)$} {
		// Test all classes
		
		\ForEach{$ltc \in LTC_c$} {
			Execute $ltc$
		}

		// Test all properties:

		\ForEach{Global Property \textbf{gp} such that $has(c, gp)$} {
			\ForEach{Global Test Property such that \textbf{gtp} $\in GTP_{gp}$} {
				\ForEach{Local Test Property such that \textbf{ltp} $\in LTP_{gtp}$} {
					Execute $ltp$
				}
			}
		}
	}
}

\caption{Executing the tests by using the metamodel (ExecTests).}
\label{alg:executionTests}
\end{algorithm}

\section{Example}
\label{sec:example}

	This section contains a very basic example showing the strengths of the proposed metamodel in a real OOP environment like Java.

\subsection{First example: Cow, grass and mouse}

\begin{figure}
\centering
\includegraphics[scale=0.43]{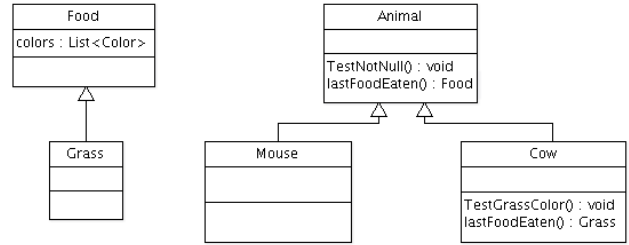}
\caption{A very simple example: Cow, grass and mouse.}
\label{fig:example1}
\end{figure}

	The example (Fig. \ref{fig:example1}) contains a super class \textit{Animal} and two subclasses \textit{Cow} and \textit{Mouse}. The \textit{Animal} class defines a method $lastFoodEaten$ which returns the last food that the animal has recently eaten. That class also introduces a new test property $TestNotNull$ associated to the method $lastFoodEaten$. This test verify that the food exists when the animal has recently eaten something. The class \textit{Cow} redefines the method lastFoodEaten and adds a new test $TestGrassColor$ in order to verify that the grass has the right color (Green).

\begin{figure}
\centering
\includegraphics[scale=0.28]{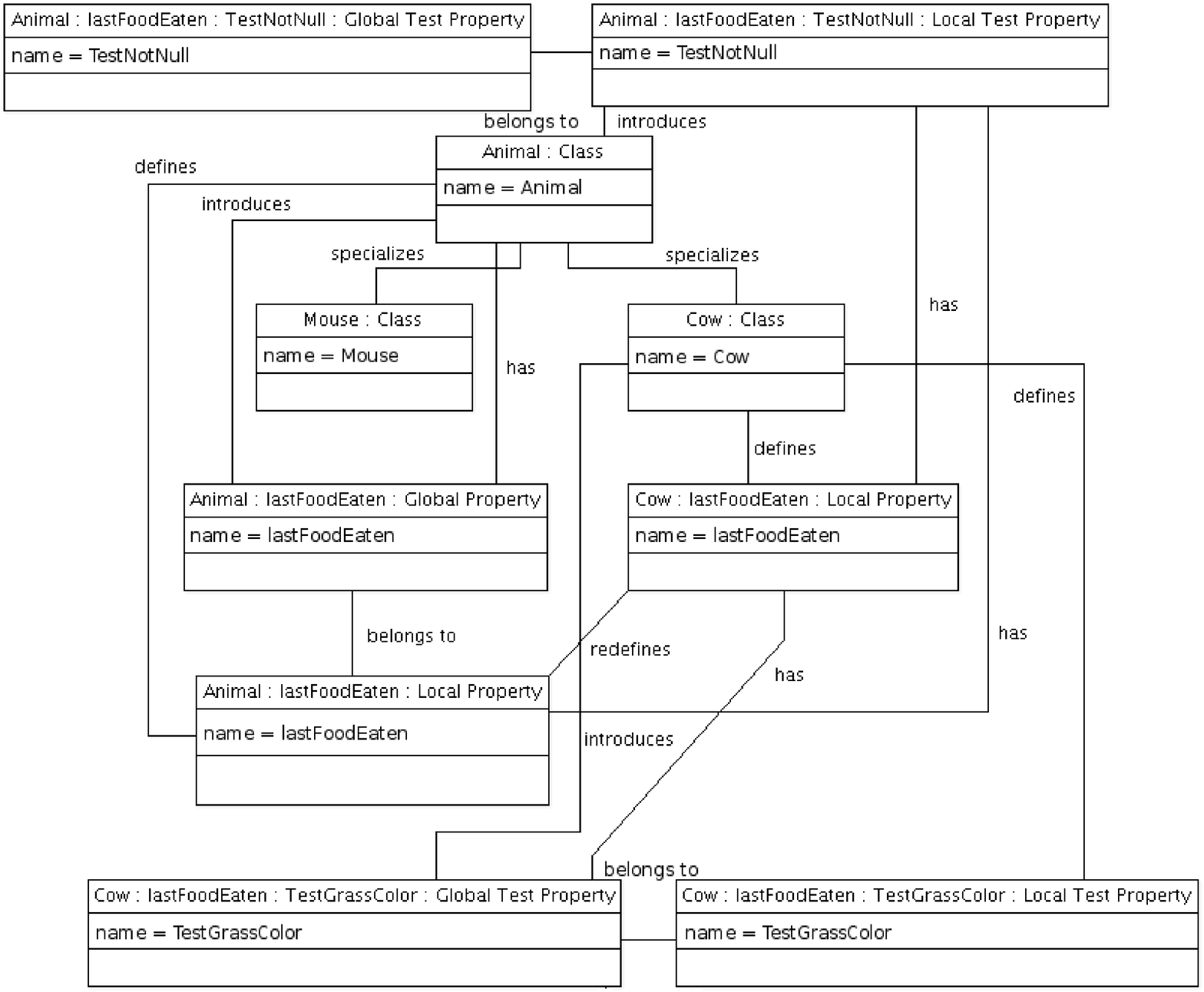}
\caption{Metamodel of the example for the classes Animal, Mouse and Cow.}
\label{fig:metaModelExample1}
\end{figure}

	The following listing presents the pseudocode of the first example:

\begin{lstlisting}[caption={Pseudocode of the example.}, label={lst:exempleAnimaux}, numbers=left]
public class Food 
{
	private List<Color> colors = 
		new ArrayList<Color>();

	public Food(Color c)
	{
		colors.add(c);
	}

	public Food()
	{
	}

	// Returns true if the food is c.
	public boolean isColor(Color c) 
	{ /* ... */ } 
}

public class Grass extends Food
{
	public Grass()
	{
		colors.add(Color.Green);
	}
}

public class Animal 
{
	private Food lastFoodEaten = null;

	Test TestNotNull 
	{
		Current.lastFoodEaten = new Food(Color.Red);

		Food last = Current.lastFoodEaten();
		assertTrue(last != null);

		String output = 
			this.getClass().getName() + ".TestNotNull";
		System.out.println(output);
	} 
	public Food lastFoodEaten()
	{
		String output = 
			this.getClass().getName() + 
			".lastFoodEaten() v1"
		System.out.println(output);

		return lastFoodEaten;
	}
}

public class Mouse extends Animal
{
}

public class Cow extends Animal
{
	Test TestGrassColor 
	{
		// Ok, it is grass.
		Current.lastFoodEaten = new Grass();

		Food f = Current.lastFoodEaten();
		assertTrue(f.isColor(Color.Green)); 
		System.out.println(
			this.getClass().getName() + 
			".TestGrassColor");
	} 
	public Grass lastFoodEaten()
	{
		System.out.println(
			this.getClass().getName() + 
			".lastFoodEaten() v2");

		if (lastFoodEaten instanceof Grass)
			return (Grass)lastFoodEaten;
		
		return null;
	}
}
\end{lstlisting}

	The listing \ref{lst:exempleAnimaux} contains a new keyword \textit{Current} which correspond to a new instance of the current class, allocated only during the test. The following syntax:

\begin{lstlisting}
Test TestName { ... } 
f()
\end{lstlisting}

	defines a Property Test named $TestName$ associated to the method $f$. Let say that a compiler tests those classes in the order of the declaration. Then, the following output would be printed:

\begin{itemize}
\item Animal.lastFoodEaten() v1
\item Animal.TestNotNull
\item Mouse.lastFoodEaten() v1
\item Mouse.TestNotNull
\item Cow.lastFoodEaten() v2
\item Cow.TestNotNull
\item Cow.lastFoodEaten() v2
\item Cow.TestGrassColor
\end{itemize}

	We can clearly observe the benefit of using the proposed metamodel: the tests defined in a class are also used in the subclasses in order to cover more lines of code automatically.

\section{Conclusion and Future Work}
\label{sec:conclusion}

	There exist two approaches in order to test a OOP software: specification-based and program-based. The approach based on a specification enables the programmer to modelize the behavior of the objects, but this technique is not so interesting since it uses an extra language and the mechanisms used does not take into account the advantages of an OOP language.

	The proposed metamodel offers a new approach combining the advantages of the current approaches. The metamodel makes it possible to write unit tests inside the classes to be tested. It facilitates the programmer to write and maintain a software unit. The proposed metamodel contains test properties for the class properties, for the classes and for the packages. There exists specialization links between the tests properties. The metamodel enables the compiler to execute automatically the tests.

	The future work is to embed the proposed metamodel in a real OOP environment in order to validate the usability and the reliability. Then, existing systems using for example JUnit will be converted to a system using the proposed metamodel in order to compare the testing approaches.

\bibliographystyle{plain}
\bibliography{icse2010}

\end{document}